\documentclass[10pt,aps,prd,twocolumn,showpacs,superscriptaddress,eqsecnum,longbibliography,nofootinbib]{revtex4-1}
\usepackage{amsmath,amsfonts,amssymb,amsthm,mathrsfs,latexsym,epsfig,enumerate,times}

\usepackage{xcolor}
\definecolor{navy}{RGB}{0,0,150}
\usepackage[colorlinks, linkcolor=navy, citecolor=navy, urlcolor=navy, plainpages=false, pdfstartview=FitH]{hyperref}
\usepackage{appendix}
\allowdisplaybreaks

\def\be{\begin{equation}}
\def\ee{\end{equation}}
\def\ba{\begin{eqnarray}}
\def\ea{\end{eqnarray}}
\def\nn{\nonumber}

\newcommand{\mubar}{{\bar \mu}} 
\newcommand{\abs}[1]{{\left|{#1}\right|}} 
\newcommand{\ket}[1]{\vert{#1}\rangle} 
\newcommand{\bra}[1]{\langle{#1}\vert} 
\newcommand{\kt}{{\tilde{K}}} 
\newcommand{\ct}{\tilde{c}}
\newcommand{\ints}{{\int_\Sigma}} 

\newcommand{\sgn}{\mathrm{sgn}} 
\newcommand{\Tr}{\mathrm{Tr}} 
\newcommand{\grav}{\mathrm{gr}} 
\newcommand{\sca}{\mathrm{sc}} 
\newcommand{\kin}{\mathrm{kin}} 
\newcommand{\hil}{\mathcal{H}} 

\allowdisplaybreaks

\begin{document}

\title{Loop quantum cosmological model from ADM Hamiltonian}
\author{You Ding}
\affiliation{School of Science, Beijing Jiaotong University, Beijing, 100044, China}
\author{Xiangdong Zhang}
\thanks{Corresponding author}
\email{scxdzhang@scut.edu.cn}
\affiliation{Department of Physics, South China University of Technology, GuangZhou 510641, China}


\begin{abstract}
The loop quantum cosmological model from ADM Hamiltonian is studied in this paper. We consider the spatially flat homogeneous FRW model. It turns out that the modified Friedmann equation keeps the same form as the APS LQC model. However, the critical matter density for the bounce point is only a quarter of the previous APS model, i.e. $\rho_{cL}=\frac{\rho_c}{4}$. This is interesting because the lower critical bounce density means the quantum gravity effects will get involved earlier than the previous LQC model. Besides, the lower critical density also means the detection of quantum gravity effects easier than the previous model.

\end{abstract}


\maketitle

\section{Introduction}\label{sec:introduction}

Loop quantum gravity (LQG) is a quantum gravity model which trying to quantize the Einstein's general relativity (GR) by using background independent techniques. LQG has been widely
investigated in last decades\cite{Ro04,Th07,As04,Ma07}.
Recently, LQG method has
been successfully generalized from GR to the metric $f(R)$ theories\cite{Zh11,Zh11b},
Brans-Dicke theory \cite{ZM12a} as well as scalar-tensor
theories of gravity \cite{ZM11c}.
However, due to the extreme complexity
of the full theory of LQG, one approach usually
taken to bypass this difficulty is to study some simpler
symmetry-reduced models. Though these symmetry-reduced models looks relatively simple, it still captures some useful ingredients of the full theory of LQG, and therefore could be used to test the constructions of LQG and to draw some physically meaningful predictions. One famous of such symmetry-reduced model from LQG is the so-called loop quantum cosmology (LQC). We refer to \cite{LQC5,Boj,APS3,AS11}
for reviews on LQC.

Just like in any quantization procedure of a classical theory, different regularization schemes exist also in LQC as well as in LQG \cite{Th07,As04,Lewandowski15}. In particular, for the LQC model of flat Friedmann-Lemaitre-Robertson-Walker (FLRW) universe, alternative Hamiltonian constraint operators were proposed \cite{APS,YDM}. In the recently proposed model, different with the Ashtekar-Pawlowski-Singh (APS) model\cite{APS}, one treats the so-called Euclidean term and Lorentzianian term of the Hamiltonian constraint independently \cite{YDM,Pawlowski18}. It was shown in \cite{Pawlowski18,Pawlowski19} that this model can lead to a new de
Sitter epoch evolution scenario where the prebounce
geometry could be described at the effective level. Then a natural question arise is that apart these two existed LQC model, are there any other possible Hamiltonian operator which can lead to a evolution different with the existed LQC model? Therefore, this paper is aimed to explore such possibility.

Note that in standard LQC, particularly in the homogeneous and spatially flat $k=0$ models, the Euclidean term and the Lorentzian term are proportional to each other.
Hence in the famous APS model of LQC, one only quantize the Euclidean term, and resulting
to a symmetric bounce \cite{APS}. However, this quantization scheme is not the only option, although it is popular in the current literature. An alternative option different with the existed model is to let  the classical theory being the well-known Arnowitt-Deser-Misner (ADM) Hamiltonian and then quantized it. It is well-known that the classically equivalent expressions would generally be nonequivalent after quantization. In particular, given the fact that the quantization expression evolved with Euclidean term and the Lorentzian term looks quite differently, it hardly to believe the resulted quantum evolution will be exactly the same as APS model. And this paper is devoted to the detailed investigation of LQC model with purely Lorentzian term and compare it with the well known APS model.

This paper is organized as follows:
After the short introduction, we give the Hamiltonian constraint we used in this paper and derive the classical evolution equations of the universe in the section \ref{sec:2}. Then
we construct the corresponding cosmological kenimatics in section
\ref{section3}, where the dynamical difference equation which representing evolution of the universe is also derived. In section \ref{section4},
bounce behaviour and effective equations are derived in section \ref{section6}. The conclusion and
some outlook are also presented in the last section.

\section{An Alternative Hamiltonian constraint in loop quantum gravity}\label{sec:2}

The Hamiltonian formulation of GR, is defined on the spacetime
manifold $M$ which could be foliated as $M=R\times \Sigma$, where $\Sigma$ is being
a three-dimensional spatial manifold and $R$ is a real line represents the time variable. The classical phase space of LQG consists of the so-called Ashtekar-Barbero variables $(A^a_i,E^i_a)$ \cite{Th07}, where $A^a_i$ is an $SU(2)$ connection and $E^i_a$ an orthonormal triad with densityweight one. The non-vanishing Poisson
bracket is given by
\ba
\{A^a_i(x),E^j_b(y)\}=8\pi G\gamma\delta^a_b\delta^j_i\delta(x,y),
\ea where $G$ is the gravitational constant and $\gamma$ the Barbero-Immirzi parameter \cite{Th07}.

The classical dynamics of GR is encoded to the three
constraints on this phase space, including the Gaussian, the diffeomorphism and the
Hamiltonian constraint. In homogeneous $k=0$ models of cosmology, the Gaussian
and the diffeomorphism constraints are automatically satisfied. Then we only need to
consider the remaining Hamiltonian constraint.

The Hamiltonian constraint in the full theory of LQG reads \cite{Th07}\cite{Lewandowski15}
\ba
H_{g}=\frac{1}{2\kappa}\ints d^3xN\left[F^j_{ab}-(\gamma^2+1)\varepsilon_{jmn}K^m_aK^n_b\right]\frac{\varepsilon_{jkl}
E^a_kE^b_l}{\sqrt{q}},\nn\\\label{Hamiltoniang}
\ea where $\kappa=8\pi G$ $N$ is the lapse function, $q$ denotes the determinant  of the spatial metric, $F^i_{ab}=\partial_aA^i_b-\partial_bA^i_a-2\varepsilon^i_{jk}A^j_aA^k_b$ and $K^i_a$ represents
the extrinsic curvature of the spatial manifold $\Sigma$. The so-called Euclidean term $H^E$ and Lorentzian term $H^L$ in Eq. \eqref{Hamiltoniang} are denoted respectively as
\ba
H^E&=&\frac{1}{16\pi G}\ints d^3 x N F^j_{ab}\frac{\varepsilon_{jkl}
E^a_kE^b_l}{\sqrt{q}},\label{HamiltonianE}
\ea and
\ba
H^L&=&\frac{1}{16\pi G}\ints d^3xN\left(\varepsilon_{jmn}K^m_aK^n_b\right)\frac{\varepsilon_{jkl}
E^a_kE^b_l}{\sqrt{q}}.\label{HamiltonianL}
\ea
Note that the famous ADM Hamiltonian reads \ba
H_{ADM}=\frac{1}{16\pi G}\int_M d^3x\sqrt{q}\left(K^{ab}K_{ab}-K^2-{}^3R\right)\label{HADM}
\ea with $K_{ab}$ and ${}^3R$ being the extrinsic curvature and curvature scalar of spatial slice $M$. Inspired by the ADM Hamiltonian, by using the relation $K^i_a=K_{ab}e^b_i$ with $q^{ab}=\delta^{ij}e^a_ie^b_j$. A direct calculation shows
\ba
H_{ADM}&=&-\frac{1}{2\kappa}\int d^3x\left[\varepsilon_{jmn}K^m_aK^n_b\frac{\varepsilon_{jkl}
E^a_kE^b_l}{\sqrt{q}}+\sqrt{q}R\right].\nn\\\label{HamiltoniancL}
\ea
Here the relation between $q_{ab}$ and the variable $E_a^i$
is given by $q_{ab} = E_a^i E_b^i /| \mathrm{det}E|$. Moreover, $\kappa=8\pi G$ and the lapse function is fixed as $N=1$  for
homogeneous universe in the current paper. Note that the APS model of LQC is only evolve the Euclidean term as\ba
H_{APS}=-\frac{1}{2\kappa\gamma^2}\int d^3 x F^j_{ab}\frac{\varepsilon_{jkl}
E^a_kE^b_l}{\sqrt{q}}.
\ea While the Hamiltonian constraint \eqref{HADM} does not contain the Euclidean term, we call this form of Hamiltonian constraint as \emph{purely Lorentzian}. We start from this form.

Now we consider the
homogeneous and isotropic $k=0$ model.
According to the cosmological principle, the metric
Friedman-Robertson-Walker (FRW) universe reads
\ba
ds^2=-dt^2+a^2(t)\left(dr^2+r^2(d\theta^2+\sin^2\theta d\phi^2)\right) \nn\ea
where $a(t)$ is the scale factor. At the classical level, one assumes that the universe be filled by some perfect fluid with matter density $\rho$ and pressure $P$.

Moreover, we introduce a massless scalar field $\phi$ as the matter content of the universe, we denote the conjugate momenta of the scalar field as $\pi$, and the commutator between them reads
\ba
\{\phi(x),\pi(y)\}&=&\delta(x,y)
\ea

In order to mimic the full theory of LQG, we do the following symmetric reduction procedures of the connection formalism as in standard LQC. First, we introduce an ``elemental cell" $\mathcal {V}$ on the spatial manifold $\mathbb{R}^3$ and restricts all
integrals to this elemental cell. Then, we choose a fiducial Euclidean metric ${}^oq_{ab}$ on $\mathbb{R}^3$ which equipped with the orthonormal triad and co-triad
$({}^oe^a_i ; {}^o\omega^i_a)$, such that
${}^oq_{ab}={}^o\omega^i_a{}^o\omega^i_b$.
For simplicity, the volume of the elemental cell $\mathcal {V}$ is measured by
${}^oq_{ab}$ and denote as $V_o$. For $k=0$ FRW model we also have $A_a^i=\gamma
\kt_a^i$, where $\gamma$ is a nonzero real number and $\kt_a^i$ is defined in \cite{ZM11c}. By fixing the local gauge and
diffeomorphism degrees of freedom, the reduced connection and densitized
triad can be obtained as \cite{LQC5}
\ba A_a^i=\ct
V_0^{-\frac13}\ {}^o\omega^i_a,\quad\quad\quad
E^b_j=pV_0^{-\frac23}\sqrt{\det({}^0q)}\ {}^oe^b_j, \nn\ea
where $\abs{p}=a^2V_0^{\frac 23}$ and
$\ct=\gamma\dot{a} V_0^{\frac 13}$ \cite{LQC5}. Hence the phase space of
cosmological model consists of conjugate pairs $(\ct,p)$ and
$(\phi,\pi)$. The non-vanishing Poisson brackets between them read
\ba
\{\ct,p\}&=&\frac{\kappa}{3}\gamma,\nn\\
\{\phi,\pi\}&=&1. \label{poissonb}\ea

The Gaussian and diffeomorphism constraints are vanished in the $k=0$ model. Hence, the
remaining Hamiltonian constraint (\ref{HamiltoniancL}) reduces to
\ba
H_G&=&H_{ADM}+H_{matter}=-\frac{3\ct^2\sqrt{\abs{p}}}{\gamma^2\kappa}+\frac{p^2_\phi}{2\abs{p}^{\frac32}}=0.\nn\\\label{Chamilton}\ea
The equation of motion of geometrical variable $p$ reads
\ba
\dot{p}=\{p,H_G\}=2\frac{\ct\sqrt{\abs{p}}}{\gamma}.
\ea
Then the classical Friedmann equation is
\ba
H^2&=&\left(\frac{\dot{a}}{a}\right)^2=\left(\frac{\dot{p}}{2p}\right)^2\nn\\
&=&\frac{\ct^2}{\gamma^2 p},
\ea
where $H$ is the Hubble parameter. By using the Hamiltonian constraint \eqref{Chamilton}, we found that
\ba
H^2=&=&\frac{\kappa}{3}\frac{p^2_\phi}{2\abs{p}^{3}}=\frac{\kappa}{3}\rho
\ea
where the matter density $\rho=\frac{p^2_\phi}{2V^2}=\frac{p^2_\phi}{2\abs{p}^{3}}$.

\section{Kinematic structure of Loop Quantization cosmology} \label{section3}

To quantize the cosmological model, we first need to construct the
corresponding quantum kinematics of cosmology by the so-called polymer-like quantization. The
kinematical Hilbert space for the geometry part can be defined as
$\mathcal{H}_{\kin}^{\grav}:=L^2(R_{Bohr},d\mu_{H})$, where
$R_{Bohr}$ and $d\mu_{H}$ are the Bohr
compactification of the real line and Haar measure on it
respectively \cite{LQC5}. Moreover, we choose
the standard Schrodinger representation for the scalar field \cite{AS11}. Thus
the kinematical Hilbert space for the scalar field part is defined as
in usual quantum mechanics,
$\mathcal{H}_{\kin}^{\sca}:=L^2(R,d\mu)$. Hence the whole Hilbert
space of LQC model is a direct product, $\hil_\kin=\hil^\grav_\kin\otimes
\hil^\sca_\kin$. Now let $\ket{\mu}$ being the eigenstates of
 $\hat{p}$ operator in the kinematical Hilbert space $\mathcal{H}_{\kin}^{\grav}$ such that
 \ba
\hat{p}\ket{\mu}=\frac{8\pi G\gamma\hbar}{6}\mu\ket{\mu}. \nn\ea
Then those eigenstates satisfy orthonormal condition
\ba
\bra{\mu_i}{\mu_j}\rangle=\delta_{\mu_i,\mu_j}\ , \ea
where $\delta_{\mu_i,\mu_j}$ is the Kronecker delta function.
For the convenience of studying quantum dynamics, we introdue new
variables
\ba v:=2\sqrt{3}sgn(p)\mubar^{-3},\quad b:=\mubar \ct, \nn\ea
where
$\mubar=\sqrt{\frac{\Delta}{|p|}}$ with
$\Delta=4\sqrt{3}\pi\gamma{\ell}_{\textrm{p}}^2$ being a minimum nonzero
area of LQG \cite{Ash-view}. They also
form a pair of conjugate variables as
\ba \{b,v\}=\frac{2}{\hbar}\ .\nn
\ea
It turns out that the eigenstates of
 $\hat{v}$ also contribute an orthonormal basis in $\mathcal{H}_{\kin}^{\grav}$.
We denote
$\ket{\phi,v}$ as the generalized orthonormal basis for the whole Hilbert space
$\hil_\kin$.

\subsection{Hamiltonian constraint of LQC with purely Lorentzian term}

Notice that the spatial curvature $R$ is vanished in the $k=0$ homogenous cosmology, the Hamiltonian constraint (\ref{HamiltoniancL}) reduces to
\ba
H_{ADM}&=&\frac{1}{2\kappa}\int d^3x\left[\varepsilon_{jmn}K^m_aK^n_b\right]\frac{\varepsilon_{jkl}
E^a_kE^b_l}{\sqrt{q}}\label{Hamiltoniana}
\ea which is purely Lorentzian term. Note that there is no operator existed corresponding to the connection variable $A^i_a(x)$ in LQG. Hence, the curvature $F^j_{ab}$ in \eqref{HamiltoniancL} should be expressed through holonomies. This can be accomplished by using Thiemann's tricks as
\ba
F^k_{ab}=-2\lim_{\mu\rightarrow 0}\Tr\left(\frac{h^{(\mu)}_{ij}\tau^k}{\mu^2}\right){}^o\omega^i_a{}^o\omega^j_b
\ea
where $h^{(\mu)}_{ij}=h^{(\mu)}_{i}h^{(\mu)}_{j}(h^{(\mu)}_{i})^{-1}h^{(\mu)}_{j})^{-1}$ is the holonomy around the loop formed by the two edges of $V$ that are
tangent to $e^a_i$
and $e^b_j$
with length $\mu$. Moreover, we also have \ba
\frac{\varepsilon^{jkl}
E^a_kE^b_l}{\sqrt{q}}=\lim_{\mu\rightarrow 0}\frac{2\sgn(p)\Tr(h^{(\mu)}_{i}\{(h^{(\mu)}_{i})^{-1},V\}\tau^j)}{\kappa\gamma\lambda}{}^o\omega^i_a\varepsilon^{abc}\nn\\
\ea Next to deal with the Lorentzian term, we also need the following identities \ba
\tilde{K}=\ints d^3x\tilde{K}^i_aE^a_i=\frac{1}{\gamma^2}\{H^E(1),V\},
\ea and\ba
\tilde{K}^m_a=\frac{1}{2\kappa^2\gamma^3}\{A_a^m,\{H^E(1),V\}\}
\ea where $H^E(1)$ is the Euclidean term and $V$ denotes the volume \cite{Th07}.

With these ingredients, the Hamiltonian constraint can be written as
\ba
H_{ADM}=\lim_{\mu\rightarrow 0}H^{\mu}
\ea with
\begin{widetext}
\ba
H^{\mu}&=&-\frac{\sgn(p)}{\kappa^5\gamma^7\mu^3}\varepsilon^{ijk}\Tr\left(h^{(\mu)}_{i}\{(h^{(\mu)}_{i})^{-1},V\}h^{(\mu)}_{j}\{(h^{(\mu)}_{j})^{-1},V\}\{(h^{(\mu)}_{k})^{-1},V\}\right)\nn\\
\label{Hamiltonian1}
\ea
\end{widetext}

The action of this operator on a quantum state $\Psi(v,\phi)$ is already known in the literature \cite{YDM}. The result is a difference equation.
Hence the final result is
\begin{widetext}
\ba
\hat{H}_{ADM}\Psi(v,\phi)&=&g_+(v)\Psi(v+8,\phi)+g_0(v)\Psi(v,\phi)+g_-(v)\Psi(v-8,\phi),
\ea
where
\ba
g_+(v)&=-\frac{\sqrt{6}}{2^{8}\times 3^3}\,\frac{\gamma^{3/2}}{\kappa^{3/2}\hbar^{1/2}}\,\frac{1}{L}\Big[M_v(1,5)f_+(v+1)-
     M_v(-1,3)f_+(v-1)\Big]\nonumber\\
       &\quad\times(v+4)M_v(3,5)\nonumber\\
       &\quad\times\Big[M_v(5,9)f_+(v+5)-M_v(3,7)f_+(v+3)\Big],\nonumber\\
g_-(v)&=-\frac{\sqrt{6}}{2^{8}\times 3^3}\,\frac{\gamma^{3/2}}{\kappa^{3/2}\hbar^{1/2}}\,\frac{1}{L}\Big[M_v(1,-3)f_-(v+1)-M_v(-1,-5)f_-(v-1)\Big]\nonumber\\
       &\quad\times(v-4)M_v(-5,-3)\nonumber\\
       &\quad\times\big[M_v(-3,-7)f_-(v-3)-M_v(-5,-9)f_-(v-5)\big],\nonumber\\
g_o(v)&=-\frac{\sqrt{6}}{2^{8}\times 3^3}\,\frac{\gamma^{3/2}}{\kappa^{3/2}\hbar^{1/2}}\,\frac{1}{L}\Big[M_v(1,5)f_+(v+1)-M_v(-1,3)f_+(v-1)\Big]\nonumber\\
       &\quad\times(v+4)M_v(3,5)\nonumber\\
       &\quad\times\Big[M_v(5,1)f_-(v+5)-M_v(3,-1)f_-(v+3)\Big]\nonumber\\
       &+\Big[M_v(1,-3)f_-(v+1)-M_v(-1,-5)f_-(v-1)\Big]\nonumber\\
       &\quad\times(v-4)M_v(-5,-3)\nonumber\\
       &\quad\times\Big[M_v(-3,1)f_+(v-3)-M_v(-5,-1)f_+(v-5)\big],
\ea
\end{widetext}
where
\ba
M_v(a,b):=|v+a|-|v+b|.
\ea

Thus, the Hamiltonian constraint (\ref{Hamiltonian1}) has been successfully quantized in the cosmological setting. The resulted Hamiltonian constraint equation of LQC turns out to be
\ba
\left(\hat{H}_{ADM}+\frac{\sqrt{3}\hat{p}_\varphi^2}{(\Delta)^{\frac32}}\widehat{\abs{v}^{-1}}\right)\Psi(\phi,v)=0.\label{hbd}
\ea
Note that in the quantum theory, the whole
Hilbert space consists of a direct product of two parts as
$\hil_\kin^{total}=\hil^\grav_\kin\otimes\hil^{matter}_\kin$. Then, the action of the matter field on a quantum state
$\Psi(v,\phi)\in\hil_\kin^{total}$ reads
\ba
\frac{\sqrt{3}\hat{p}_\varphi^2}{(\Delta)^{\frac32}}\widehat{\abs{v}^{-1}}\Psi(v,\phi)=-\frac{\sqrt{3}}{(\Delta)^{\frac32}}\hbar^2B(v)
\frac{\partial^2\Psi(v,\phi)}{\partial\phi^2}.\label{hm} \ea where $\widehat{\abs{v}^{-1}}\Psi(v,\phi)=B(v)\Psi(v,\phi)$ and the detailed expression of $B(v)$ can be found in \cite{APS}.

\section{Effective Hamiltonian of LQC }\label{section4}

Now we come to
study the effective
theory of this new LQC. Since we also want to know
the effect of matter fields on the dynamical evolution. Hence we
include a scalar matter field $\varphi$ into LQC. Note that the cosmological expectation value for Lorentzian term has already obtained in literature as\cite{YDM,DL18}
\ba
\langle \hat{H}^L\rangle&=\frac{3\beta}{4\gamma^2\kappa\Delta}\sin^2(2b).
\ea Then the effective total Hamiltonian
constraint \eqref{Hamiltoniana} reads
\ba
H_F&=&\langle \hat{H}_G\rangle=-\frac{3\beta}{\gamma^2\kappa\Delta}|v|\sin^2b\left(1-\sin^2(b)\right)+\beta|v|\rho\label{effH},
\ea
where $\beta=\frac{\kappa\hbar\gamma\sqrt{\Delta}}{4}$.

\section{Effective equations and the quantum bounce}\label{section6}
Now we discuss the effective dynamics.
By employing the effective Hamiltonian \eqref{effH}.
The equation of motion for $v$ reads
\ba
\dot{v}=\{v,H_F\}=\frac{6}{\hbar\gamma^2\kappa\Delta}|v|\sin(2b)\left(1-2\sin^2(b)\right)
\ea Note the bounce take place at the minimum of volume $v$, and therefore happened at the point of $\sin^2(b)=\frac12$.

So, the density can be expressed as
\ba
\rho=\frac{3}{\gamma^2\kappa\Delta}\sin^2b\left(1-\sin^2(b)\right)\leq\frac{3}{4\gamma^2\kappa\Delta}=\frac14\rho_c=\rho_{cL}\label{matterdensity}
\ea where $\rho_c=\frac{3}{\gamma^2\kappa\Delta}$ is the critical matter density in the standard LQC. By using the expression of $\dot{v}$, the modified Friedman equation can be obtained
\ba
H^2=\left(\frac{\dot{v}}{3v}\right)^2=\frac{8\pi G}{3}\rho\left(1-\frac{\rho}{\rho_{cL}}\right)\label{FriedmanADM}.
\ea Now, in order to calculate the evolution of the physical quantity such as matter density and volume of the universe, we first introduce $x=\sin^2(b)$. Consider $(x')^2$ with prime being derivative
with respect to $\phi$. From the definition of $x$, we have\ba
(x')^2=\left(2\sin(b)\cos(b)b'\right)^2=4x(1-x)b'^2,\label{xprime}
\ea and \ba
b'&=&\frac{db}{dt}\frac{dt}{d\phi}=\{b,H_F\}\frac{V}{p_\phi}=-\sqrt{12\pi G x(1-x)}
\ea Plugging the above expression into \eqref{xprime}, we find the
equation\ba
x'=2\sqrt{12\pi G}x(1-x).
\ea Solution to this equation reads\ba
x=\frac{1}{1+e^{-2\sqrt{12\pi G}\phi}}
\ea and hence from Eq. \eqref{matterdensity} \ba
\rho(\phi)=\frac{3}{\kappa\gamma^2\Delta}\frac{1}{4\cosh^2(\sqrt{12\pi G}\phi)},
\ea so the volume is \ba
V(\phi)=\frac{p_\phi}{\sqrt{2\rho}}=\sqrt{\frac{16\pi G\gamma^2\Delta p^2_\phi}{3}}\cosh(\sqrt{12\pi G}\phi).
\ea \begin{figure}[!htb]
		\includegraphics [width=0.50\textwidth]{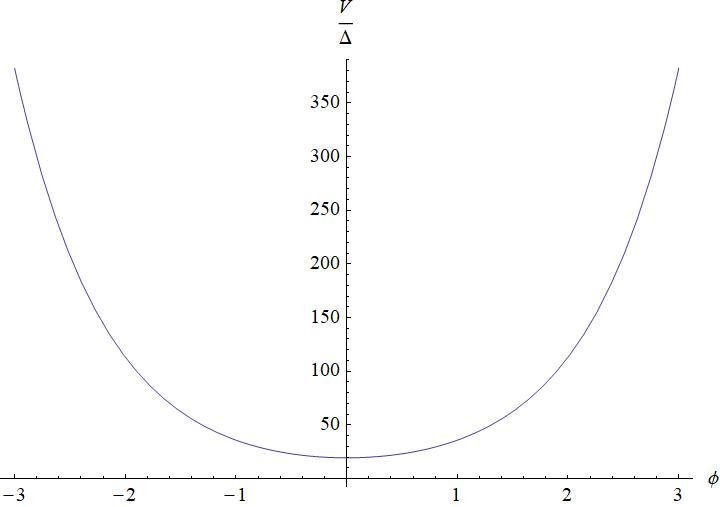}
		\caption{The $V/\Delta$ as a function of $\phi$. The $\gamma=0.2375$ and $8\pi G=1$ are adapted in the calculation.}
		\label{fig:Vofphi}
	\end{figure}

Now let us study the asymptotic behavior of the above LQC model in the classical region, namely the large $v$ region.
For $v\rightarrow\infty$ limit, the matter density $\rho$ in Eq. \eqref{matterdensity} goes to
zero, which leads to
\ba
b\rightarrow 0\quad or \quad b\rightarrow \arcsin{(1)}=\frac{\pi}{2}
\ea
in situation $b\mapsto 0$, the asymptotic of
Hamiltonian constraint read
\ba
H_F=-\frac{3\beta}{\gamma^2\kappa\Delta}|v|b^2+\beta|v|\rho.
\ea
while $b\mapsto \frac{\pi}{4}$, the asymptotic of
Hamiltonian constraint approaches to
\ba
H_F=-\frac{3\beta}{\gamma^2\kappa\Delta}|v|\left((b-\frac{\pi}{2})^2\right)+\beta|v|\rho.
\ea
Then the resulted Friedman equations read
\ba
H^2&=&\frac{\kappa}{3}\rho, \quad\quad ( b\rightarrow 0)\\
H^2&=&\frac{\kappa}{3}\rho, \quad\quad ( b\rightarrow \frac{\pi}{2})
\ea
which is also a symmetric bounce.

\section{concluding remarks}\label{section7}

To summarize, the loop quantum cosmological model consists of purely Lorentzian term is studied in this paper. We consider the spatially flat homogeneous FRW model. It turns out that the modified Friedmann equation keeps the same form as APS LQC model. However, the critical matter density for the bounce point is only a quarter of the previous APS model, i.e. $\rho_{cL}=\frac{\rho_c}{4}$. This is interesting because from Eq. \eqref{FriedmanADM} we can see that the strength of quantum gravity becoming significant when $\frac{\rho}{\rho_{cL}}\sim1$, since $\rho_{cL}<\rho_c$, in this sense the lower critical bounce density means the quantum gravity effects will get involved earlier than previous LQC model. Besides, the lower critical density also means detection of quantum gravity effects easier than previous model. It should be note that in this paper, we only consider the cosmological implication of LQC from ADM Hamiltonian which only contain purely Lorentzian term. However, since the Lorentzian term and the Euclidean term lead to different results at quantum mechanical level, then one can also natural considering the mixture of these two terms. This of course possible, the authors in \cite{Pawlowski18,Pawlowski19,Zhang21} shows that when we considering the Lorentzian term and the Euclidean term appearing in the Hamiltonian constraint simultaneously, an effective cosmological constant could emerged at the large volume limit.

It should be noted that there are many aspects of the new LQC which deserve further investigating.
For examples, it is still desirable to perturbation theory of the new LQC, in this case, the spatial curvature will not be zero. And thus could be inherent more features from the full theory of LQG.  Moreover, the authors in \cite{Yang19} adapt an alternative regularization procedure via Chern-Simons theory which is quite different with the usual regularization method in LQG, and the resulted cosmological evolution is different with APS LQC model. Hence it also interesting to study this regularization under our framework of new LQC. We leave all these interesting
topics for future study.

\begin{acknowledgements}
This work is supported by NSFC with Grants No. 11775082.

\end{acknowledgements}



\begin{thebibliography}{99}



\bibitem{Ro04} C. Rovelli, {\it Quantum Gravity,} (Cambridge University Press, 2004).

\bibitem{Th07} T. Thiemann, {\it Modern Canonical Quantum General Relativity,} (Cambridge University
Press, 2007).


\bibitem{As04}A. Ashtekar and J. Lewandowski, {\it Background independent quantum gravity: A
status report,} Class.Quant.Grav. {\bf21}, R53 (2004).

\bibitem{Ma07} M. Han, W. Huang, and Y. Ma, {\it Fundamental structure of loop quantum gravity,}
 Int. J. Mod. Phys. D {\bf16}, 1397 ,(2007).


\bibitem{Zh11} X. Zhang and Y. Ma, {\it Extension of loop quantum gravity to $f(R)$
theories,} Phys. Rev. Lett. {\bf 106}, 171301 (2011).


\bibitem{Zh11b} X. Zhang and Y. Ma, {\it Loop quantum f(R) theories,} Phys. Rev. D {\bf 84}, 064040 (2011).


\bibitem{ZM12a} X. Zhang and Y. Ma, {\it Loop quantum Brans-Dicke
theory,} J. Phys.: Conf. Ser. {\bf 360}, 012055 (2012).


\bibitem{ZM11c} X. Zhang and Y. Ma,  {\it Nonperturbative loop
quantization of scalar-tensor theories of gravity,} Phys. Rev. D {\bf 84}, 104045 (2011).


\bibitem{LQC5}
A. Ashtekar, M. Bojowald, and J. Lewandowski, {\it Mathematical structure of loop quantum cosmology,} Adv. Theor. Math.
Phys. \textbf{7}, 233 (2003).

\bibitem{Boj}
M. Bojowald, {\it Loop quantum cosmology,} Living Rev. Relativity \textbf{8}, 11 (2005).

\bibitem{AS11} A. Ashtekar, P. Singh, {\it Loop quantum cosmology: A status
report,}  Class. Quant. Grav. {\bf28}, 213001 (2011).

\bibitem{APS3} A. Ashtekar, T. Pawlowski, P. Singh,  {\it Quantum nature of the big bang: Improved
dynamics,}  Phys. Rev. D {\bf 74}, 084003 (2006).


\bibitem{Ash-view}A.~Ashtekar, {\it Loop quantum cosmology: An overvie,} Gen. Rel. Grav. {\bf41}, 707 (2009).

\bibitem{ACS}A. Ashtekar, A. Corichi, and P. Singh, {\it Robustness of key features of loop quantum
cosmology,} Phys. Rev. D  {\bf 77}, 024046 (2008).

\bibitem{Taveras}
V. Taveras, {\it Corrections to the Friedmann equations from loop quantum gravity for a universe with a free scalar field,} Phys. Rev. D \textbf{78}, 064072 (2008).

\bibitem{DMY}
Y. Ding, Y. Ma and J. Yang, {\it Effective scenario of loop quantum cosmology,} Phys. Rev. Lett. \textbf{102}, 051301
(2009).

\bibitem{YDM}
J. Yang, Y. Ding and Y. Ma, {\it Alternative quantization of the Hamiltonian in loop quantum cosmology,} Phys. Lett. B \textbf{682}, 1 (2009).

\bibitem{Boj11}
M. Bojowald, D. Brizuela, H. H. Hernandez, M. J. Koop, H. A.
Morales-Tecotl, {\it High-order quantum back-reaction and quantum cosmology with a positive cosmological constant,} Phys. Rev. D \textbf{84}, 043514 (2011).

\bibitem{ACH102}
A. Ashtekar, M. Campiglia, A. Henderson, {\it Loop quantum cosmology and spin foams,} Phys. Lett. B \textbf{681}, 347 (2009); {\it Casting loop quantum cosmology in the spin foam paradigm,} Class. Quant. Grav. \textbf{27}, 135020 (2010); {\it Path integrals and the WKB approximation in loop quantum cosmolog,} Phys. Rev. D \textbf{82}, 124043 (2010).


\bibitem{QM1}
L. Qin and Y. Ma, {\it Coherent state functional integrals in quantum cosmology,} Phys. Rev. D \textbf{85}, 063515 (2012).


\bibitem{Lewandowski15}M. Assanioussi, J. Lewandowski, I. Makinen, {\it New scalar constraint operator for loop quantum gravity,} Phys. Rev. D 92, 044042 (2015).


\bibitem{Pawlowski18}M. Assanioussi, A. Dapor, K. Liegener, and T. Pawlowski, {\it Emergent de sitter epoch of the quantum cosmos
from loop quantum cosmology}, Physical Review Letters, 121(8):081303, 2018.


\bibitem{Pawlowski19}M. Assanioussi, A. Dapor, K. Liegener, and T. Pawlowski, {\it Emergent de Sitter epoch of the Loop Quantum Cosmos: a detailed analysis},  Phys. Rev. D 100, 084003 (2019).

\bibitem{APS}A. Ashtekar, T. Pawlowski, and P. Singh, {\it Quantum Nature of the Big Bang}, Phys. Rev. Lett. 96, 141301 (2006).

\bibitem{YDM}
J. Yang, Y. Ding and Y. Ma, {\it Alternative quantization of the Hamiltonian in loop quantum cosmology,} Phys. Lett. B \textbf{682}, 1 (2009).

\bibitem{DL18}A. Dapor, K. Liegener, {\it Cosmological Effective Hamiltonian from full Loop Quantum Gravity Dynamics,} Phys. Lett. B785, 506 (2018),
\bibitem{Zhang21}X. Zhang, G. Long, Y. Ma, {\it Loop quantum gravity and cosmological constant,} Physics Letters B 823, (2021) 136770.


\bibitem{Yang19}J. Yang, C. Zhang, Y. Ma, {\it Loop quantum cosmology from an alternative Hamiltonian,} Phys. Rev. D 100, 064026 (2019).









\end{thebibliography}
\end{document}